\documentclass{article}

\usepackage[preprint]{neurips_2022}

\usepackage[utf8]{inputenc} 
\usepackage[T1]{fontenc}    
\usepackage{url}            
\usepackage{booktabs}       
\usepackage{amsfonts}       
\usepackage{nicefrac}       
\usepackage{microtype}      
\usepackage{xcolor}         

\usepackage{enumitem}
\makeatletter
\newcommand{\printfnsymbol}[1]{%
  \textsuperscript{\@fnsymbol{#1}}%
}
\usepackage{algorithmicx,algpseudocode,algorithm}
\usepackage{wrapfig,tikz}
\usepackage{amsmath,longtable,fancyhdr,booktabs,multirow,graphicx,float}
\usepackage{adjustbox, bigstrut, tabularx, multirow, makecell, diagbox}
\usepackage{amssymb,xcolor,amsthm}
\usepackage{subcaption}

\newcommand{\be}{\begin{equation}}
	\newcommand{\ee}{\end{equation}}

\definecolor{Gray}{gray}{0.85}
\definecolor{LightCyan}{rgb}{0.88,1,1}

\newcolumntype{a}{>{\columncolor{Gray}}c}

\makeatletter
\usepackage{xspace}
\def\@onedot{\ifx\@let@token.\else.\null\fi\xspace}
\DeclareRobustCommand\onedot{\futurelet\@let@token\@onedot}

\newcommand{\figref}[1]{Fig\onedot~\ref{#1}}

\title{Brain PET Synthesis from MRI Using Joint Probability Distribution of Diffusion Model at Ultrahigh Fields}

\author{%
    Taofeng Xie\thanks{TaoFeng Xie and Chentao Cao contributed equally to this work.}\\
    Inner Mongolia University \\
    Inner Mongolia Medical University \\
    \texttt{tf.xie@mail.imu.edu.cn} 
    \And
    Chentao Cao\printfnsymbol{1}  \\ 
    SIAT, Chinese Academy of Sciences \\ 
    \texttt{ct.cao@siat.ac.cn} 
    \AND
    Zhuoxu Cui  \\ 
    SIAT, Chinese Academy of Sciences \\ 
    \texttt{zx.cui@siat.ac.cn} 
    \And
    Fanshi Li  \\ 
    SIAT, Chinese Academy of Sciences \\ 
    \texttt{fs.li@siat.ac.cn}
    \And
    Zidong Wei  \\ 
    SIAT, Chinese Academy of Sciences \\ 
    \texttt{zd.wei@siat.ac.cn}
    \And
    Yanjie zhu \\ 
    SIAT, Chinese Academy of Sciences \\ 
    \texttt{yj.zhu@siat.ac.cn}
    \And
    Ye Li  \\ 
    SIAT, Chinese Academy of Sciences \\ 
    \texttt{liye1@siat.ac.cn}
    \And
    Dong Liang  \\ 
    SIAT, Chinese Academy of Sciences \\ 
    \texttt{dong.liang@siat.ac.cn}
    \And
    Qiyu Jin  \\ 
    Inner Mongolia University \\ 
    \texttt{qyjin2015@aliyun.com}
    \And
    Guoqing Chen   \\
    Inner Mongolia University \\ 
    \texttt{cgq@imu.edu.cn}
    \And
    Haifeng Wang\thanks{Corresponding author}  \\ 
    SIAT, Chinese Academy of Sciences \\ 
    \texttt{hf.wang@siat.ac.cn} 
}

\begin{document}
\maketitle
\begin{abstract}
MRI and PET are important modalities and can provide complementary information for the diagnosis of brain diseases because MRI can provide structural information of brain and PET can obtain functional information of brain. However, due to the expensive expense of PET scanning or radioactive exposure, some patients do not accept it, resulting in a lack of PET scans. Especially, simultaneous PET and MRI imaging at ultrahigh field is not achievable in the current. Thus, synthetic PET using MRI at ultrahigh field is essential. In this paper, we synthetic PET using MRI as a guide by joint probability distribution of diffusion model (JPDDM). On the public the Alzheimer’s Disease Neuroimaging Initiative (ADNI) dataset, we contrasted CycleGAN and score-based of SDE, and our model achieved a great result. Meanwhile, We utilized our model in 5T MRI and 7T MRI.
\end{abstract}

\section{Introduction}
Diagnosing the disease of brain disorder, (e.g. Alzheimer’s disease (AD)) jointed Positron emission tomography (PET) and magnetic resonance imaging (MRI) become a popular and useful method because they can offer various information \cite{johnson2012brain, zhang2017pet, cheng2017cnns}. MRI and PET include complementing information for enhancing the accuracy of AD diagnosis \cite{calhoun2016multimodal, liu2017multi}. 
Positron emission tomography (PET) uses radiolabeled molecules like 18F-fluorodeoxyglucose (FDG) to offer metabolic information. MRI scan can provide structural information. However, the expensive expense of PET scanning or radioactive exposure, many patients will to receive MRI scans and to reject PET scans. Ultrahigh field MRI (e.g.5T MRI and 7T MRI) can offers images which is higher resolution and high signal-to-noise ratio. But the PET corresponding to ultrahigh field cannot be obtained. Missing PET scans be synthesized urgently in order to make effective use of the multi-modality medical data, especially for PET at ultrahigh field. 
Generative models have generative adversarial networks (GAN) \cite{goodfellow2020generative}, likelihood-based method \cite{graves2013generating} and diffusion model \cite{sohl2015deep} and so on. 
In recent years, diffusion model have remarkable advances comparable to GANs \cite{goodfellow2020generative}. 

Score matching with Langevin dynamics (SMLD) \cite{song2019generative} calculates the score namely the gradient of the log probability density with respect to the data at each noise scale and sample from a series of diminishing noise scales in the process of generation by Langevin dynamics. Denoising diffusion probabilistic modeling (DDPM) \cite{ho2020denoising} uses functional knowledge of the reverse distributions to make training tractable. It trains a series of probabilistic models to reverse each stage of the noise corruption. The model of score-based generative solving the stochastic differential equation (SDE) \cite{song2020score} unified framework of SMLD \cite{song2019generative} and DDPM \cite{ho2020denoising}. SDE diffuses a data point into random noise continuously and then reverse the process for molding random noise into raw data. This paper's core task is synthesizing PET scans from MRI scans. Therefore, we need to estimate the probability distribution of PET ($x_{PET}$) conditioned on MRI ($x_{MRI}$), i.e., $p(x_{PET}|x_{MRI})$. However, it is not easy to estimate. In the framework of diffusion models described above, we only need to estimate $\nabla_{x_{PET}}p(x_{PET}|x_{MRI})$. Given $\nabla_{x_{PET}}p(x_{PET}|x_{MRI})=\nabla_{x_{PET}}p(x_{PET}, x_{MRI})$, we will construct an MRI and PET joint diffusion model and learn its joint distribution to achieve conditional generation.
\section{Method}
\begin{figure}[H]
    \centering
    \includegraphics[width=1\textwidth]{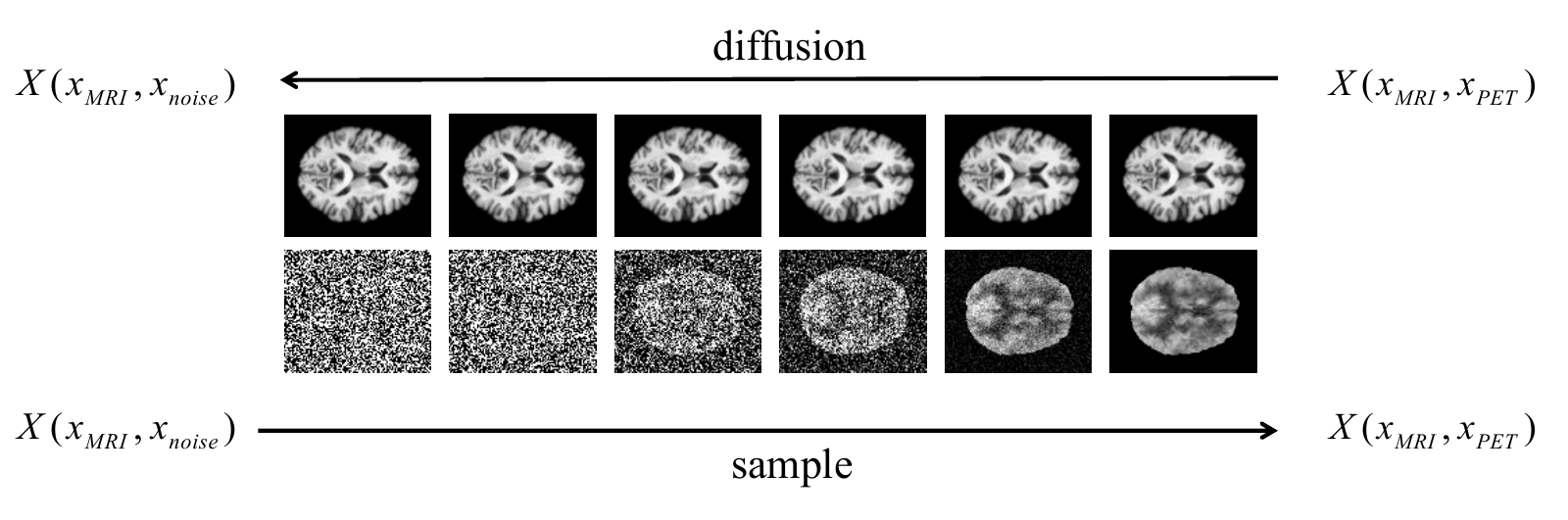}
    \caption{schematic diagram of joint distribution in this work}
    \label{fig:schematic diagram}
\end{figure}

Diffusion modeling is a crucial method in the generating process. The method forecasts the score, namely the gradient of the log probability density with respect to original data, not the data distribution directly. The purpose of this paper is to synthetic PET by MRI as a guidance. We provided appropriate conditions as a guidance for the model in order to get the expected results. The joint probability distribution of diffusion model (JPDDM) has two processes that are diffusion process and sample process.  \figref{fig:schematic diagram}is schematic diagram of joint distribution in this study. 
The diffusion process is forward SDE. The diffusion process is as follows
\begin{equation*}
X_{i+1} = X_{i} + \sigma_{min} \left( \frac{\sigma_{max}}{\sigma_{min}} \right) ^{t} z, i=1,2,...,N-1,
\end{equation*} 
 $X_{i}$ is the $i-th$ joint perturbed data (e.g. $X_{i}(x_{PET},x_{MRI},t)$). $X_{0}$ is joint distribution of $x_{PET}$ and $x_{MRI}$. $x_{T}$ obeys joint distribution of Standard Gaussian distribution and $x_{MRI}$. $\{\sigma_{i}\}_{i=1}^{N}$ is the noise scales which $\sigma_{min}$ is the minimum of the noise scales and $\sigma_{max}$ is the maximum of the noise scales.
Sample process is predictor-corrector sample namely PC sample. Predictor and corrector are executed alternately. Predictor is reverse diffusion ( from joint distribution of noise and MRI to the distribution of PET) for the sample that can be described as
\begin{equation*}
X_{i} = X_{i+1} - f_{i+1}(X_{i+1}) + g_{i+1}(X_{i+1})g_{i+1}(X_{i+1})^{T} s_{\theta^{*}} \left (X_{i+1},i+1 \right) + g_{i+1}(X_{i+1})z_{i+1}
\end{equation*}
Where $f_{i}$ denotes the drift coefficient of $X_{i}$. $g_{i}$ denotes the diffusion coefficient of $X_{i}$. $s_{\theta^{*}(X_{i+1},i+1)}$ is to estimate $\nabla_{X_{i}}\log p_{t}(X_{i})$. $p_{t}(X_{i})$ is the distribution of $X_{i}$. $s_{\theta^{*}}(X_{i+1},i+1)$ is obtained by deep learning of UNet and objection function is


\begin{equation*}
L(\theta; \sigma)=\frac{1}{2}\mathbb{E}_{p_{t}(X)}\left[\left\| \sigma_{min} \left(\frac{\sigma_{max}}{\sigma_{min}}\right)^{t} s_{\theta}(X_{i+1},\sigma)+z\right\|_{2}^{2} \right]
\end{equation*}
In this study, $f_{i}=0$, $g_{i}=\sqrt{\sigma_{i}^{2}-\sigma_{i-1}^{2}}$.
Corrector is Langevin dynamics. Langevin method can computes the sample by
\begin{equation*}
X_{i} = X_{i+1} + \varepsilon s_{\theta}(X, i+1) + \sqrt{2\varepsilon} z
\end{equation*}
Where $\varepsilon=2 \alpha_{i}\left(r\lVert z\rVert_{2} / \lVert s_{\theta}\rVert_{2}\right)$ denotes step size.
The study utilized the Alzheimer’s Disease Neuroimaging Initiative (ADNI) dataset \cite{jack2008alzheimer}. 14440 pairs image of MRI and PET had registered. All image are reshaped to 128 * 128.
\section{Results}
\begin{figure}[H]
    \centering
    \includegraphics[width=1\textwidth]{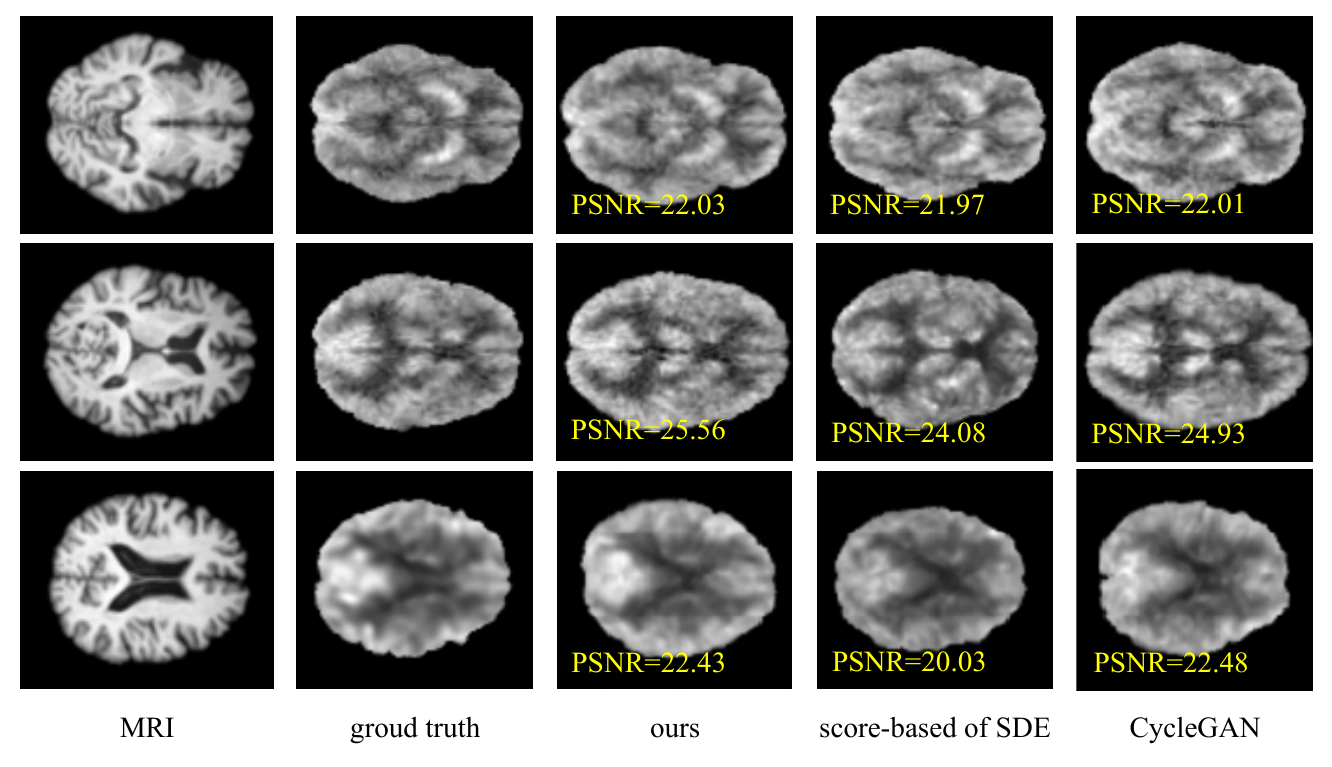}
    \caption{comparison synthetic results using different generation methods(joint distribution, CycleGAN, score-based of SDE)
}
    \label{fig:comparison results}
\end{figure}

\begin{figure}[H]
    \centering
    \includegraphics[width=1\textwidth]{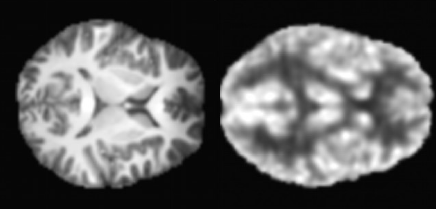}
    \caption{Synthetic PET using 5T MRI. The first picture is 5T MRI, and the second picture is synthetic PET using the first picture}
    \label{fig:5T}
\end{figure}

\begin{figure}[H]
    \centering
    \includegraphics[width=1\textwidth]{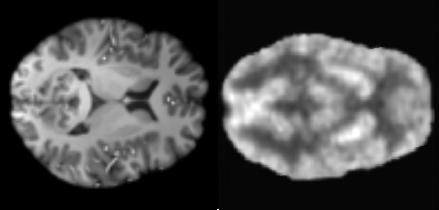}
    \caption{Synthetic PET using 7T MRI. The first picture is 7T MRI, and the second picture is synthetic PET using the first picture}
    \label{fig:7T}
\end{figure}

In this study, Peak Signal to Noise Ratio (PSNR) were used to assess image quality. Our model, CycleGAN \cite{zhu2017unpaired} and score-based of SDE model \cite{song2020score} are contrasted for synthesis PET using MRI in our study. The experimental results are shown in \figref{fig:comparison results}. 
Efficiency of our model is better and PSNR of our model is hig
her than others. We applied the trained model to the 5T MRI images acquired by a 5T MRI scanner (uMR Jupiter, United Imaging, Shanghai, China) and the 7T MRI images acquired by a 7T MRI scanner (MAGNETOM Terra, Siemens Healthcare, Erlangen, Germany). All of the protocols were approved by our Institutional Reviews Board (IRB). The results of 5T MRI show in \figref{fig:5T}. The results of 7T MRI show in \figref{fig:7T}.
\section{Conclusions and Discussion}
This study synthetic PET from MRI using joint probability distribution of diffusion model. It not only improves the stability of the generation model but also enables more accurate recovery of PET from MRI. The method has high potential for cross-modal synthesis. However, the disadvantage of our method is the slow of imaging speed. In future research, accelerated imaging speed is one of the research directions.


\clearpage
\bibliography{refs}




\end{document}